# Tunable superconductivity in electron- and hole-doped Bernal bilayer graphene

Chushan Li[1,2], Fan Xu[1], Bohao Li[3], Jiayi Li[1], Guoan Li[4], Kenji Watanabe[5], Takashi Taniguchi[6], Bingbing Tong[4], Jie Shen[4], Li Lu[4,7], Jinfeng Jia[1,2,7,8], Fengcheng Wu[3,9]*, Xiaoxue Liu[1,2,7,8]*, and Tingxin Li[1,2,7]*

[1]Key Laboratory of Artificial Structures and Quantum Control (Ministry of Education), School of Physics and Astronomy, Shanghai Jiao Tong University, Shanghai 200240, China

[2]Tsung-Dao Lee Institute, Shanghai Jiao Tong University, Shanghai, 201210, China

[3]School of Physics and Technology, Wuhan University, Wuhan 430072, China

[4]Beijing National Laboratory for Condensed Matter Physics and Institute of Physics, Chinese Academy of Sciences, Beijing 100190, China

[5]Research Center for Electronic and Optical Materials, National Institute for Materials Science, 1-1 Namiki, Tsukuba 305-0044, Japan

[6]Research Center for Materials Nanoarchitectonics, National Institute for Materials Science, 1-1 Namiki, Tsukuba 305-0044, Japan

[7]Hefei National Laboratory, Hefei 230088, China

[8]Shanghai Research Center for Quantum Sciences, Shanghai 201315, China

[9]Wuhan Institute of Quantum Technology, Wuhan 430206, China

*Emails: wufcheng@whu.edu.cn, xxliu90@sjtu.edu.cn, txli89@sjtu.edu.cn

**Graphene-based, high quality two-dimensional electronic systems have emerged as a highly tunable platform for studying superconductivity[1-21]. Specifically, superconductivity has been observed in both electron-doped and hole-doped twisted graphene moiré systems[1-17], whereas in crystalline graphene systems, superconductivity has so far only been observed in hole-doped rhombohedral trilayer[18] and hole-doped Bernal bilayer graphene (BBG)[19-21]. Recently, enhanced superconductivity has been demonstrated[20,21] in BBG due to the proximity with a monolayer $WSe_2$. Here, we report the observation of superconductivity and a series of flavor-symmetry-breaking phases in both electron- and hole-doped BBG/$WSe_2$ device by electrostatic doping. The strength of the observed superconductivity is tunable by applied vertical electric fields. The maximum Berezinskii–Kosterlitz−Thouless (BKT) transition temperature for the electron- and hole-doped superconductivity is about 210 mK and 400 mK, respectively. Superconductivities emerge only when applied electric fields drive BBG electron or hole wavefunctions toward the $WSe_2$ layer, underscoring the importance of the $WSe_2$ layer in the**

**observed superconductivity. We find the hole-doped superconductivity violates the Pauli paramagnetic limit, consistent with an Ising-like superconductor. In contrast, the electron-doped superconductivity obeys the Pauli limit, even though the proximity induced Ising spin-orbit coupling is also notable in the conduction band. Our findings highlight the rich physics associated with the conduction band in BBG, paving the way for further studies into the superconducting mechanisms of crystalline graphene and the development of novel superconductor devices based on BBG.**

**Introduction**

Although intrinsic spin-orbit-coupling (SOC) effects are negligible in graphene, SOC can be induced through the proximity effect by directly contacting graphene with transition metal dichalcogenide layers[22-25]. Experimentally, such van der Waals SOC proximity method has been demonstrated as an important tuning knob for engineering the physical properties of graphene-based systems[13,14,20,21,26-34]. For example, the proximity-induced Ising SOC is considered as a key factor in stabilizing the superconducting state in BBG/WSe$_2$ heterostructures[20,21]. However, the specific pairing mechanisms of superconductivity in both graphene moiré systems and crystalline graphene systems are still an ongoing research topic[35-37]. On the other hand, in crystalline graphene, although the interaction-driven flavor-symmetry-breaking phases have been observed in both conduction band (CB) and valence band (VB)[38-45], superconductivity has so far only been observed in the VB[18-21]. Here we report the observation of tunable superconductivity in BBG/WSe$_2$ system. Benefiting from the high vertical electrical displacement field $D$ achievable in the device, the electron-doped superconductivity is observed for the first time in crystalline graphene.

**Electron- and hole-doped Superconductivities**

The geometry of BBG/WSe$_2$ device is shown in Fig. 1a, where the dual gates of $V_{top}$ and $V_{back}$ allow for independent control of $D$ and carrier density $n$ in BBG (Methods). Figure 1d shows the longitudinal resistance $R_{xx}$ as a function of $D$ and $n$ at zero magnetic field, covering both the electron- and hole-doped regions. In the measured $D$ and $n$ ranges, the WSe$_2$ layer is kept as charge neutral due to the type-I band alignment shown in Fig. 1b (Methods). A series of phase transitions featured with peaks or dips in $R_{xx}$ can be observed in the $R_{xx}$-$D$-$n$ map. Figure 1e illustrates the corresponding phase diagram based on Fig. 1d and Fermi surface analysis via quantum oscillations. Notably, the measured $R_{xx}$ exhibits clear electron-hole asymmetry and $D$-field asymmetry. The electron-hole asymmetry is associated with the asymmetric band structure of the CB and VB in BBG (Fig. 1c). The $D$-field asymmetry is related to the fact that the proximity induced Ising SOC effect is only significant on the top graphene layer which is closer to the WSe$_2$ layer[20,21,23,24,31]. In our device, at positive (negative) $D$, hole wavefunctions concentrate at the top (bottom) layer of the BBG, and electron wavefunctions concentrate at the bottom (top) layer of the BBG. Therefore, notable spin splitting is either in the CB ($D < 0$) or in the VB (for $D > 0$), as illustrated in Fig. 1c. Experimentally, the estimated Ising SOC strength $\lambda_I$ is around 1.7 meV in our device (Extended Data Fig. 1).

Remarkably, two notable superconducting regions at finite doping emerge in the *n-D* phase diagram, one is at positive *D* with hole doping, and another is at negative *D* with electron doping. Both regions correspond to the case that doped carriers are polarized to the top graphene layer. The hole-doped superconductivity initiates at about 0.85 V/nm, which is consistent with the previous observations[20,21], while the electron-doped superconductivity initiates at a negative *D* ~ -1.25 V/nm. Both the electron and hole superconducting states move to the higher doping density with increasing |*D*|. The strength of the superconductivity, characterized by the critical current $I_c$ and the critical temperature $T_c$, can be effectively tuned by applied *D*. Figure 1f and 1i show the differential resistance ($dV_{xx}/dI_{dc}$) versus dc bias current ($I_{dc}$) of the hole- and electron-doped superconducting states at various *D*, respectively. On the hole-doped side, both $I_c$ and $T_c$ show nonmonotonic dependence on *D* (Fig. 1g). At the optimal *D* and *n* of the hole-doped superconductivity, $T_c$ (determined by 50% of the normal state resistance) is about 450 mK (Fig. 1h). Moreover, $T_{BKT}$ is estimated to be about 400 mK by fitting the nonlinear *I-V* traces (Extended Data Fig. 2), which is about 1.7 times higher than previous studies[20,21]. Additionally, the critical current densities are significantly larger than previous reports[20,21] (Methods). On the electron-doped side, within the achievable range of *D*, $I_c$ and $T_c$ increase with increasing of |*D*| but exhibit a saturating trend (Fig. 1j). The maximum $T_c$ and $T_{BKT}$ is about 300 mK and 210 mK, respectively (Fig. 1k).

In crystalline graphene systems, the emergence of flavor-symmetry-breaking phases and superconductivity is considered to be associated with the van Hove singularities (VHS), characterized by divergent single-particle density of states (DOS), near the band edge. Extended Data Fig. 3 illustrates the calculated single-particle DOS of BBG at various displacement fields. First of all, both the CB and VB exhibit clear VHS, but at a given *D*, the VHS in the CB locates at lower carrier densities. At relatively small *D*-fields, the VHS is more pronounced in the VB, whereas at larger *D*, the VHS becomes stronger in the CB. These results are consistent with our observations that the flavor-symmetry-breaking phases emerge at lower *n* and larger *D* on the electron-doped side. Secondly, although the carrier density corresponding to the VHS increases monotonically with increasing *D*, the magnitude of DOS near VHS reach its maximum at certain *D* values. Since the effective Coulomb interactions are reduced at higher carrier density due to screening, the dependence of VHS on *D* shown in Extended Data Fig. 3 indicates that there is an optimized *D*-field range for engineering correlation physics in BBG. This aligns with the experimental observation that both superconductivities and symmetry-breaking states appear within a specific *D*-field range.

**Quantum oscillations on the hole-doped side**

Investigating quantum oscillations can provide important information of Fermi surface structures. Figure 2a displays the measured $R_{xx}$ as a function of perpendicular magnetic field $B_\perp$ and $n$ at $D = 1.1$ V/nm on the hole-doping side. Several sets of quantum oscillations can be distinguished through Fast Fourier transform (FFT) of $R_{xx}$ ($1/B_\perp$), as shown in Fig. 2c. The frequency $f_v$ is normalized by the total carrier density, representing the fraction of the total Fermi surface area enclosed by a given cyclotron orbit (Methods).

The results at $D = 1.1$ V/nm near the hole-doped superconducting region closely resemble those reported in previous studies[20,21]. In short, at hole densities lower than the superconducting region, two frequencies with $f_v^{(1)} > 1/12$ and $f_v^{(2)} < 1/12$ can be identified, indicating a state with six smaller Fermi pockets and six larger Fermi pockets. In contrast, at $D = -1.1$ V/nm, the FFT frequency peak only occurs at $f_v = 1/12$ (Extended Data Fig. 4), corresponding to the spin and valley symmetric Fermi surface with 12 degeneracies resulting from trigonal warping[46]. This is consistent with the scenario that the WSe$_2$ induced Ising SOC lifts the spin and valley degeneracy of the BBG valence band at positive $D$. Inside the superconducting region, FFT analysis of quantum oscillations for the normal state shows spectral weight mostly concentrate at frequency slightly less than 1/2 and at low frequencies, corresponding to a partial isospin-polarized phase with two major Fermi pockets and multiple minor Fermi pockets (denoted as PIP$_2$ phase), consistent with the prior reports[20,21]. Interestingly, with further increasing hole doping beyond the PIP$_2$ phase, the system evolves into a state with four annular Fermi surfaces, which is evident by two FFT frequency peaks satisfying $f_v^{(1)} - f_v^{(2)} = 1/4$. The annular Fermi surface is also evident in Fig. 2a that a set of electron-like Landau fan emerges at $n \sim -1.3 \times 10^{12}$ cm$^{-2}$. Note that in RTG, superconductivity emerges from an annular Fermi sea, near a PIP$_2$ phase[18]. Here, the BBG/WSe$_2$ system exhibits similar Fermi surface conditions but lacks an observable superconducting phase in the annular Fermi sea, highlighting a significant distinction between these two systems. The results at $D = 1.19$ V/nm are essentially similar to those observed at $D = 1.1$ V/nm, while an additional $R_{xx}$ dip emerges at low densities (Extended Data Fig. 5).

When the applied $D$-field is beyond 1.4 V/nm, the hole-doped superconductivity vanishes. Simultaneously, the Fermi surface structures become less complex. Extended Data Fig. 4b and 4d display the quantum oscillations at $D = -1.5$ V/nm and its corresponding FFT, respectively. A single frequency peak at $f_v = 1/12$, resulting from trigonal warping, is observed over a wide density range at $D = -1.5$ V/nm. At $D = 1.5$ V/nm, due to the SOC-induced spin splitting in VB, the 12 degenerated Fermi pockets transform into six smaller Fermi pockets and six bigger Fermi pockets (Fig. 2d). In general, no superconductivity and flavor-symmetry-breaking states are observed at $D = \pm 1.5$ V/nm, likely due to the weakened VHS and interaction effects at such a high $D$-field, as discussed above.

**Quantum oscillations on the electron-doped side**

Figure 3a and 3b present the $R_{xx}$ as a function of $B_\perp$ and $n$ for electron doping at $D = 1.55$ V/nm and -1.55 V/nm, respectively. Their FFT results are shown in Fig. 3c and 3d correspondingly. At $D = 1.55$ V/nm and high electron densities, the Fermi surface is spin and valley symmetric with fourfold degeneracy ($f_v = 1/4$), as expected for a simplest graphene system without flavor-symmetry-breaking and trigonal-warping effects. With decreasing electron densities, spontaneous flavor symmetry breaking occurs, reducing the Fermi surface degeneracy to two-fold then finally to one-fold. These results resemble the half-metal and quarter-metal phases reported in rhombohedral trilayer[38], tetralayer[42], and pentalayer graphene[43]. Importantly, in between of the normal metal phase (4-fold degeneracy) and the half-metal phase (2-fold degeneracy), as well as in between of the half-metal phase and the quarter-metal phase (1-fold degeneracy), we observe partially isospin polarized Fermi surfaces ($PIP_1$ and $PIP_2$) similar to the hole-doped case. At $D = -1.55$ V/nm, the observed Fermi surface structures with tuning electron density are quite similar to the $D = 1.55$ V/nm case, except that SOC plays an important role at $D = -1.55$ V/nm, which can be identified from the FFT frequency peak splitting around $f_v = 1/4$ in the normal metal phase (Fig. 3d). Note that in the phase with $f_v = 1/2$, no FFT frequency peak splitting around $f_v = 1/2$ can be observed, indicating two Fermi pockets with identical area. The evolution of Fermi surface with electron doping density is qualitatively captured by our theoretical calculations (Extended Data Fig. 6). Remarkably, akin to the hole-doped superconductivity, electron-doped superconductivity again only emerges when electron wavefunctions are tuned close to the $WSe_2$ layer, and it also originates from normal states possessing a $PIP_2$ Fermi surface. In RTG, $PIP_1$ and $PIP_2$ Fermi sea have not been reported in the CB[18,38], which could provide a clue to the absence of superconductivity in electron-doped RTG. The evolution of Fermi surface with changing electron densities at $D = \pm 1.64$ V/nm (Extended Data Fig. 8) closely resemble those observed at $D = \pm 1.55$ V/nm.

At $|D| < 1.25$ V/nm, superconductivity is not observed on the electron-doping side, but the flavor-symmetry-breaking phases persist to lower $D$. Extended Data Fig. 7 shows quantum oscillations and their FFT results at $D = \pm 1.1$ V/nm. The half-metal and quarter-metal phases can be clearly identified from the FFT results. However, at $D = \pm 1.1$ V/nm, the $PIP_1$ phase is absent and the electron density range of the $PIP_2$ phase become much narrower compared to that at $D = \pm 1.64$ and $\pm 1.55$ V/nm, which is also consistent with the theoretically calculated results (Extended Data Fig. 6). These observations again indicate the importance of the PIP Femi sea for superconducting pairing in this system.

**Response to in-plane magnetic field**

Although the electron-doped and the hole-doped superconductivity in $BBG/WSe_2$ system seems have similar origin from the Fermi surface analysis, their response to in-plane magnetic field is quite different. Figure 4a and 4b show $R_{xx}$ as a function of $n$ and $T$ at $D = 0.96$ V/nm (hole-doped) and $D = -1.64$ V/nm (electron-doped), respectively. The superconducting paring strength is comparable at $D = 0.96$ V/nm on the hole-doping side and at $D = -1.64$ V/nm on the electron-doping side, as evidenced by similar values of $T_c$ and the critical perpendicular magnetic field $B_{c\perp}$ (Fig. 4c and 4d). Nevertheless, their response to the in-plane magnetic field $B_{\parallel}$ differs significantly, as shown in Fig. 4e and 4f. The hole-doped superconductivity is resilient to applied $B_{\parallel} = 1$ T within a certain range of $n$, while the electron-doped superconductivity is completely suppressed at a much lower $B_{\parallel} \sim 0.2$ T.

For two-dimensional spin-singlet superconductors, the suppression of superconductivity in the presence of external $B_{\parallel}$ is primarily attributed to Zeeman effect. This gives an upper bound of the critical in-plane magnetic field $B_{c\parallel}$, known as the Pauli paramagnetic limit $B_p$. For weakly-coupled BCS superconductors, $B_p = 1.86$ (T/K) $\times T_c^0$ with $g$-factor = 2, where $T_c^0$ is the critical temperature at zero magnetic field. Figure 4g and 4h depict the Pauli violation ratio (PVR) $B_{c\parallel}^0/B_p$ as a function of carrier density at $D$ = 0.96 V/nm and $D$ = -1.64 V/nm, respectively. The critical in-plane magnetic field at the zero-temperature limit $B_{c\parallel}^0$ is obtained by fitting the $B_{c\parallel}(T)$ versus the temperature to the phenomenological formula $T/T_c^0 = 1 - (B_{c\parallel}/B_{c\parallel}^0)^2$ (see Extended Data Fig. 9). The ratio of $B_{c\parallel}^0/B_p$ over the hole-doped superconducting dome changes from ~2.4 to ~1.7 with increasing hole doping, explicitly violating the Pauli paramagnetic limit. On the contrary, the electron-doped superconductivity obeys the Pauli paramagnetic limit with $B_{c\parallel}^0/B_p < 1$ across the whole superconducting dome.

The hole-doped superconductivity is consistent with the phenomenology of Ising superconductivity[47-49], which results from the proximity-induced Ising SOC by WSe$_2$, consistent with previous reports[20,21]. However, the limited resilience to $B_{\parallel}$ observed in electron-doped superconductivity is more puzzling, as a comparable Ising SOC effect is evident in the CB of BBG/WSe$_2$ at negative $D$ fields. This requires a different scenario than the spin-valley locking picture from typical Ising superconductors. For example, more complicated scenarios such as intervalley coherent states may need to be considered[20]. On the other hand, Rashba SOC and in-plane orbital effects could also compete with Ising SOC, leading to the observed suppression of the $B_{c\parallel}$. However, to align with the experimental observations, it is necessary for the Rashba SOC and/or in-plane orbital effects to have important differences in the electron-doped and hole-doped superconductors. Further theoretical and experimental studies are required to understand the underlying mechanism of the reduced PVR for electron-doped superconductivity in BBG/WSe$_2$ system.

**Conclusions**

Understanding the superconducting pairing mechanism in both crystalline graphene systems and twisted graphene systems remains as one of the most important and intriguing problems in condensed matter physics. Among all of graphene-based superconductors, BBG offers the simplest platform to study the mystery of the emergent superconductivity. Additionally, the structurally stable nature of BBG is a notable advantage compared to other graphene superconductors, enabling the reproducible fabrication of high-quality devices. We have revealed a rich phase diagram tuned by $n$ and $D$ for both the hole- and electron-doped BBG proximitized with a monolayer WSe$_2$. The flavor-symmetry-breaking phases emerge at large *D*-fields in BBG closely resemble those observed in rhombohedral-stacked multilayer graphene. Both the hole- and electron-doped superconductivity are associated with the emergence of PIP$_2$ Fermi surfaces and the proximity to WSe$_2$. Our results highlight the rich physics associated with the CB in BBG, revealing both the similarities and differences between the electron- and hole-doped superconductivity in BBG/WSe$_2$. The observation that electron-doped superconductivity does not behave as an Ising-like superconductor suggests that the role of WSe$_2$ in stabilizing superconductivity in BBG may not be solely attributed to Ising SOC. These observations provide substantial constraints on theoretical models for understanding the mechanism of superconductivity in crystalline graphene systems.

## Methods

### Device Fabrication

The BBG/WSe$_2$ device has a dual graphite gate geometry and is assembled by using the standard dry-transfer technique[50]. A poly (bisphenol A carbonate)/polydimethylsiloxane stamp mounted on a glass slide is used to pick up each layer. The stacking sequence from top to bottom consists of the following layers: graphite as the top gate electrode, top hexagonal boron nitride (hBN) as the top dielectric, monolayer WSe$_2$ (Commercial source, HQ graphene), graphite as the contact electrodes, BBG, bottom hBN as the bottom dielectric, and graphite as the bottom gate electrode. The entire structure is released onto a Si/SiO$_2$ substrate at 180 °C. The stack is then shaped into a Hall-bar geometry using reactive ion etching with CHF$_3$/O$_2$, and Cr/Au (2/100nm) is deposited as the metal edge contacts by electron beam evaporation. The device image is shown in Extended Data Fig. 11.

### Transport Measurements

The dual gate geometry allows us to independently tune the carrier density $\left(n = \frac{c_t V_t + c_b V_b}{e} + n_0\right)$ and the vertical electric displacement field $\left(D = \frac{c_t V_t - c_b V_b}{2\varepsilon_0} + D_0\right)$ in BBG by applying top gate voltage $V_t$ and back gate voltage $V_b$. Here, $\varepsilon_0$, $c_t$, $c_b$, $n_0$ and $D_0$ denote the vacuum permittivity, geometric capacitance of the top gate, geometric capacitance of the back gate, intrinsic doping and the build-in electric field, respectively.

Electrical transport measurements were performed in two dilution refrigerators. One is a top-loading dilution refrigerator (Oxford TLM, nominal base temperature about 15 mK) with 18 T superconducting magnet. The sample is immersed in the He$^3$-He$^4$ mixtures during the measurements. Each fridge line has a sliver epoxy filter and a RC- filter (consisting of a 470 Ω resistor and a 100-pF capacitor) at low temperature. Another one is a bottom-loading dilution refrigerator (Oxford Triton, nominal base temperature about 10 mK) with a vector superconducting magnet of 9-1-1 T, and the in-plane magnetic field dependence measurement is performed in this dilution refrigerator. For the 9-1-1 T dilution refrigerator, each fridge line has a π-filter at room temperature, consisting of two 10 nF capacitors and a 10 mH inductor; and a RC-filter at low-temperature, consisting of a 1 kΩ resistor and a 1nF capacitor. We performed the electrical transport measurements by using the standard low-frequency (< 23 Hz) lock-in (SR830 and SR860) techniques. The bias current is limited within 3 nA, to avoid sample heating, and avoid disturbing the fragile states. Voltage pre-amplifier with 100 MΩ impedance were used to improve the signal-to-noise ratio.

It is worth noting that, on the hole-doped side, we did not observe the in-plane magnetic field-induced superconductivity ($T_c \sim 30$ mK) at $D < 0$ (reported in Ref. 19) and the SC1 phase ($T_c \sim 40$ mK) at $D > 0$ reported in Ref. 21, presumably limited by the sample quality or effective electron temperature of our dilution refrigerators.

**Superconducting coherence length, mean free path, and critical current density**

The Ginzburg-Landau superconducting coherence length $\xi$ can be estimated from the relation[51] $\xi = \sqrt{\Phi_0/(2\pi B_{c\perp})}$, where $\Phi_0 = h/2e$ is the superconducting flux quantum, and $B_{c\perp}$ is the critical perpendicular magnetic field at zero temperature. The measured $B_{c\perp}$ of both electron- and hole-doped superconductivity in our device at ~20 mK (<< $T_c$) ranges from ~ 5 mT to 15 mT (Extended Data Fig. 12), yielding $\xi \approx$ 150 nm – 250 nm. This is comparable to values reported in previous studies of crystalline graphene systems[18-21].

The mean free path $l_m$ can be estimated[19] based on the onset magnetic field $B_{onset}$ of quantum oscillations by $l_m \sim 2\pi k_F l_B^2$, where $k_F$ is the Fermi wave vector, and $l_B$ is the magnetic length at $B_{onset}$. The $k_F$ of BBG can be estimated by $k_F \sim \sqrt{\pi|n|}$. The carrier density range of the observed electron- and hole doped superconductivity is around $|n|$ = 0.5-1.2×10$^{12}$ cm$^{-2}$, and the $B_{onset}$ is about 0.2-0.3 T, giving the $l_m$ about 2 μm to 5 μm. Overall, the $\xi/l_m$ ratio are smaller than 0.1 in BBG/WSe$_2$ system, suggesting that both the electron-doped and hole-doped superconductivity are in the clean limit.

Based on the measured critical current $I_c$ (Fig. 1g, 1j) and Hall bar width (~1.5 μm, see Extended Data Fig. 11), we estimate that the critical current density $J_c$ of the hole-doped superconductivity in our device ranges from ~40 nA/μm to ~110 nA/μm, with $T_c$ ranging from ~150 mK to ~450 mK. These values are significantly larger than the previously reported $J_c$ values of ~2-5 nA/μm (with $T_c$ ~200-300 mK) in hole-doped BBG/WSe$_2$ devices[20,21]. We found the $J_c$ of the electron-doped superconductivity is smaller, ranging from ~15 nA/μm to ~40 nA/μm (with $T_c$ ~150 mK to ~300 mK) in our device.

**Normalized FFT frequency $f_v$**

$f_v$ is defined as $f_v = f_{FFT} \times e/nh$, where $f_{FFT}$ is the quantum oscillation frequency (in tesla) of $R_{xx}(B_\perp)$ derived by FFT, with $n$, $h$ and $e$ denoting the total carrier density, Planck's constant and electron charge, respectively. $f_v$ represents the fraction of the total Luttinger volume enclosed by a given cyclotron orbit. The sum of $f_v^{(i)}$ with degeneracy $N_i$ should be normalized, namely $\sum_i s_i N_i f_v^{(i)} = 1$, where $s_i$ is +1 (-1) when the $i$th cyclotron orbit encloses an electron (hole) pocket for electron doping (the rule is opposite for hole doping).

**Theoretical superconducting mechanisms for BBG**

We present a brief review about theoretical superconducting mechanisms for BBG. The proposed pairing mechanisms include electron-phonon coupling[52-55], electron-electron interactions[56-60], electronic collective fluctuations in quantum critical models[61-62], etc. The enhancement of superconductivity in BBG coupled to WSe$_2$ has often been attributed to the proximity-induced Ising spin-orbit coupling either through the electron-electron interaction mechanism[56-58] or by suppressing spin fluctuations[63]. In addition, virtual tunneling between WSe$_2$ and BBG has also been proposed to enhance superconductivity[54]. A thorough review of superconductivity and correlated phases in non-twisted graphene systems can be found in Ref. 37.

**Calculations**

We use a continuum $k.p$ model to describe the low-energy band structure of BBG,

$$H_0 = \sum_{\tau=\pm} \sum_{k} \psi_\tau^\dagger(k)[h_{0,\tau}(k)s_0]\psi_\tau(k),$$

$$h_{0,\tau}(k) = \begin{pmatrix} \frac{U}{2} & v_0\Pi^\dagger & -v_4\Pi^\dagger & -v_3\Pi \\ v_0\Pi & \Delta+\frac{U}{2} & \gamma_1 & -v_4\Pi^\dagger \\ -v_4\Pi & \gamma_1 & \Delta-\frac{U}{2} & v_0\Pi^\dagger \\ -v_3\Pi^\dagger & -v_4\Pi & v_0\Pi & -\frac{U}{2} \end{pmatrix},$$

where $\tau = \pm 1$ is the valley index, $s_0$ is the identity matrix in the spin space, and $h_{0,\tau}(k)$ is a 4 × 4 matrix expressed in the space (A1, B1, A2, B2) with A1 (A2) and B1 (B2) are sublattices in bottom (top) layers. In $h_{0,\tau}(k)$, $\Pi = (\tau k_x + i k_y)$, $U$ is the interlayer potential difference between the top and bottom layers generated by an external out-of-plane displacement field, and the velocities can be parametrized as $v_i = \sqrt{3}a_0\gamma_i/2$ for $i = 0, 3, 4$ with $a_0 = 0.246$ nm being the lattice constant of monolayer graphene. We take the following parameter values from Ref. [66], $\gamma_0 = 2.61$ eV, $\gamma_1 = 361$ meV, $\gamma_3 = 283$ meV, $\gamma_4 = 138$ meV, and $\Delta = 15$ meV. The potential $U$ is related to the displacement field $D$ by $U = eDd_0/\epsilon$, where $e$ is the elementary charge, $d_0$ is the interlayer distance of BBG, and $\epsilon$ is the dielectric constant. Given the uncertainties in the value of $\epsilon$, we use the parameter $U$ in the calculation.

A monolayer WSe$_2$ adjacent to the bilayer graphene generates spin-orbit coupling (SOC) terms given by,

$$H_{SOC} = \sum_{\tau=\pm} \sum_{k} \psi_\tau^\dagger(k)[h_{SOC,\tau}(k)]\psi_\tau(k),$$

$$h_{SOC,\tau}(k) = \left[\frac{\lambda_I}{2}\tau s_z + \frac{\lambda_R}{2}(\tau\sigma_x s_y - \sigma_y s_x)\right]\frac{l_0 - l_z}{2},$$

where the Pauli matrices $s_{x,y,z}$ and $\sigma_{x,y}$ act, respectively, on the spin and sublattice space. The operator $\frac{l_0-l_z}{2}$ projects the SOC term onto the top layer graphene, where $l_0$ and $l_z$ are, respectively, identity and Pauli $z$ matrix in the layer space. The parameters $\lambda_I$ and $\lambda_R$ quantify the strength of the Ising and Rashba SOC. In our theoretical calculation, we only keep the Ising SOC term for simplicity, and take $\lambda_I = 2$ meV and $\lambda_R = 0$ meV. Detailed effects of Rashba SOC are left for future study. The single-particle band structure and DOS shown in Fig. 1b and Extended Data Fig. 3b, 3c are calculated using the Hamiltonian $H_1 = H_0 + H_{\text{SOC}}$.

We consider Coulomb interaction described by:

$$H_C = \frac{1}{2A} \sum_{k,k',q,\alpha,\beta} V(\boldsymbol{q}) \psi_\alpha^\dagger(\boldsymbol{k}) \psi_\beta^\dagger(\boldsymbol{k}') \psi_\beta(\boldsymbol{k}'-\boldsymbol{q}) \psi_\alpha(\boldsymbol{k}+\boldsymbol{q}),$$

where α and β represent layer, sublattice, spin and valley indices, and $A$ is area of the system. We use the gate-screened Coulomb potential $V(\boldsymbol{q}) = \frac{2\pi e^2}{\epsilon_r |\boldsymbol{q}|} \tanh |\boldsymbol{q}| d$, where $\epsilon_r$ is a phenomenological dielectric constant and $d$ is the gate-to-sample distance. Here we take $\epsilon_r$ as a phenomenological parameter taking into account both environmental screening from hBN and internal metallic screening. We use Hartree-Fock approximation to study the full Hamiltonian $H = H_1 + H_C = H_0 + H_{\text{SOC}} + H_C$, and compare mean-field energies of competing states, including symmetric states and symmetry-breaking states (spin and/or valley polarized states are examined). The theoretical results for electron doping are illustrated in Extended Data Fig. 6. In the calculation, we use $\epsilon_r = 20$ and $d = 20$ nm. Related theoretical studies can be found, for example, in Refs. 64 and 65.

**Band alignment between BBG and monolayer WSe$_2$**

As illustrated in Fig. 1b, the charge neutral point of BBG is deep within the monolayer WSe$_2$ semiconducting band gap (~1.8 eV). We now argue that the WSe$_2$ layer is not doped for electric field within $D=\pm 1.65$ V/nm by considering a simple model. In the presence of an electric field, the energies of valence band maximum (VBM) and conduction band minimum (CBM) in BBG are (neglecting the tiny Ising SOC), respectively, $-|U|/2$ and $+|U|/2$, where $U = eDd_0/\epsilon$. The energies of VBM and CBM in WSe$_2$ are respectively, $E_v - U_1$ and $E_c - U_1$. Here $E_v$ ($E_c$) is the energy of WSe$_2$ VBM (CBM) measured relative to the BBG charge neutrality point in the absence of the electric field. Based on the experimentally determined band offsets in WSe$_2$ and graphene systems[67-69], we estimate that $E_v = -0.6$ eV and $E_c = 1.2$ eV. The energy shift $U_1$ due to the applied electric field is given by $U_1 = eD(d_0/2 + d_1)/\epsilon$, where $d_1$ is the interlayer distance between WSe$_2$ and its adjacent graphene layer. The energy offset in VBM (CBM) between BBG and WSe$_2$ is given, respectively, by,

$$\delta E_v = -\frac{|U|}{2} - (E_v - U_1) = \begin{cases} |E_v| + \dfrac{e|D|d_1}{\epsilon}, & D > 0 \\ |E_v| - \dfrac{e|D|(d_0 + d_1)}{\epsilon}, & D < 0 \end{cases} ;$$

$$\delta E_c = (E_c - U_1) - \frac{|U|}{2} = \begin{cases} |E_c| - \dfrac{e|D|(d_0 + d_1)}{\epsilon}, & D > 0 \\ |E_c| + \dfrac{e|D|d_1}{\epsilon}, & D < 0 \end{cases}.$$

By taking $|D|$ to be the upper limit of 1.65 V/nm, $\epsilon \approx 4$, $d_0 \approx 0.34$ nm, and $d_1 \approx 0.5$ nm, we find that all the above band offsets remain at least larger than 0.25 eV. Therefore, we conclude that the WSe$_2$ layer is not doped for electric field within $D=\pm 1.65$ V/nm. Moreover, the relevant band offsets increase with increasing $|D|$ for the two regimes: (1) $D > 0$ and hole doping to the valence band of BBG; (2) $D < 0$ and electron doping to the conduction band of BBG. These two regimes, in which superconductivity has been observed, are the main focus of our work. In experiment, we also do not find any signature of carrier doped into the WSe$_2$ layer, consistent with the above analysis.

**Acknowledgement**

We thank Yangzhi Chou, Jianpeng Liu, Yang Zhang, and Fanqi Yuan for helpful discussions. This work is supported by the National Key R&D Program of China (No. 2022YFA1405400, 2022YFA1402702, 2022YFA1402404, 2019YFA0308600, 2022YFA1402401, 2020YFA0309000), the National Natural Science Foundation of China (No. 12350403, 12174249, 92265102, 12374045), the Innovation Program for Quantum Science and Technology (Grants No. 2021ZD0302600 and No. 2021ZD0302500), the Natural Science Foundation of Shanghai (No. 22ZR1430900), the Shanghai Jiao Tong University 2030 Initiative Program. X.L. acknowledges the Pujiang Talent Program 22PJ1406700. T.L. acknowledges the Yangyang Development Fund. K.W. and T.T. acknowledge support from the JSPS KAKENHI (Grants No. 21H05233 and No. 23H02052) and World Premier International Research Center Initiative (WPI), MEXT, Japan. This work was supported by the Synergetic Extreme Condition User Facility (SECUF).

**Author contributions**

T.L. and X.L. designed the experiment. C.L. and F.X. fabricated the devices. C.L., F.X. and J.L. performed the measurements with the assistance of G.L. and B. T.

X.L., C.L., T.L. and F.W. analyzed the data. B. L. and F. W. performed theoretical studies. K.W. and T.T. grew the bulk hBN crystals. T.L., X.L., and F.W. wrote the manuscript. All authors discussed the results and commented on the manuscript.

**Competing interests**

The authors declare no competing financial interests.

# Main Figures

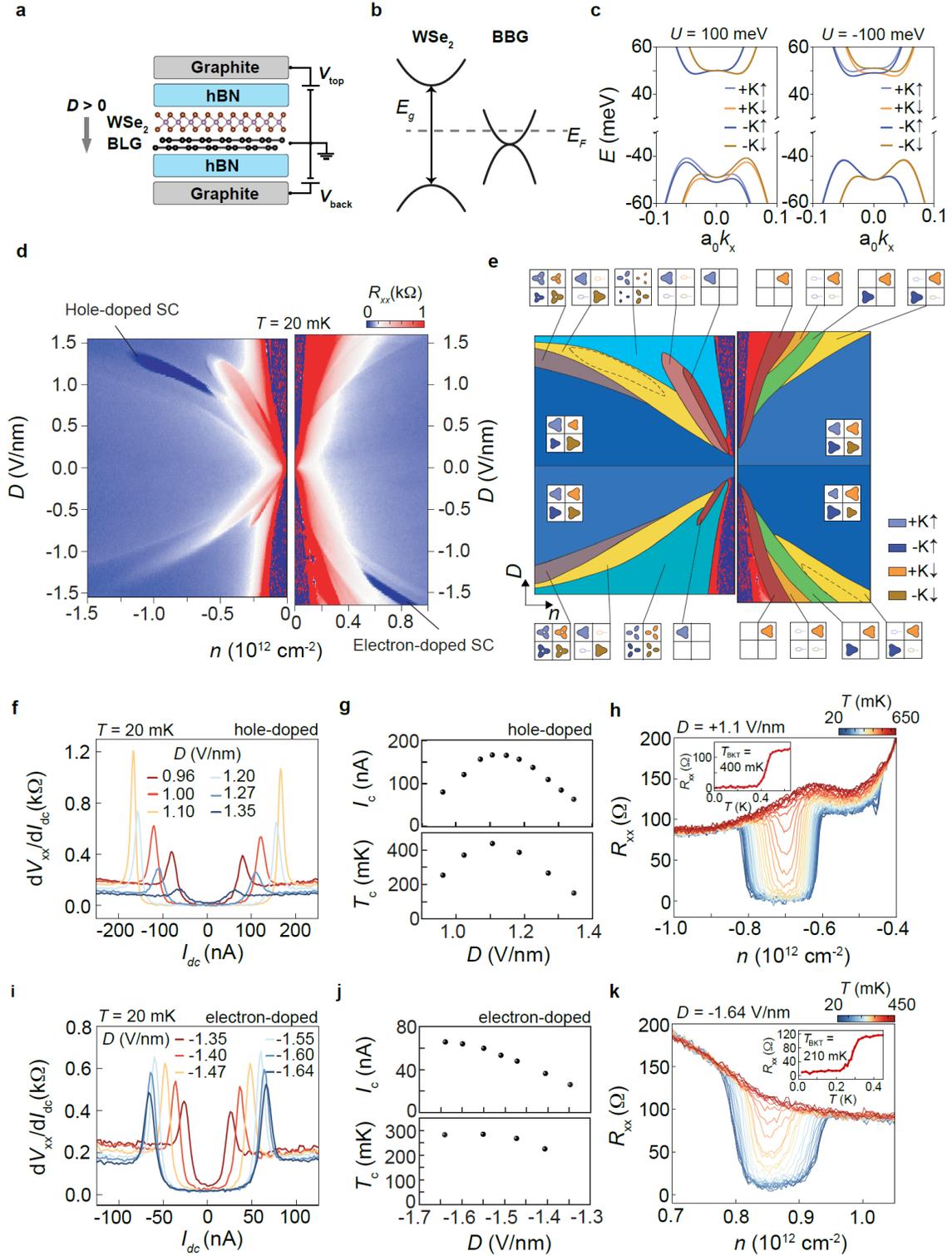

**Fig. 1. Phase diagram, electron- and hole-doped superconductivity of BBG/WSe$_2$. a**, Schematic of the dual-gated BBG/WSe$_2$ device. **b**, Schematic band alignment of monolayer WSe$_2$ and BBG at $D = 0$ V/nm. The charge neutral point of BBG is deep within the WSe$_2$ semiconducting band gap. **c**, Calculated single-particle band structure near the K and K' points of the Brillouin zone for interlayer potential difference $U = \pm 100$ meV, which roughly corresponds to the displacement field $D \approx \pm 1$ V/nm. At positive (negative) $D$, hole (electron) wavefunctions concentrate at the top layer of BBG, so the WSe$_2$ induced SOC is much more prominent in the valence (conduction) band. **d**, $R_{xx}$-$D$-$n$ map measured at $T$ = 20 mK, covering both the electron-doped ($0 < n < 1.0 \times 10^{12}$ cm$^{-2}$, -1.65 V/nm $< D <$ 1.60 V/nm) and hole-doped ($-1.5 \times 10^{12}$ cm$^{-2} < n < 0$, -1.55 V/nm $< D <$ 1.55 V/nm) regions. **e**, Experimental phase diagram determined based on **d** and FFT analysis of quantum oscillations. We use the $R_{xx}$ features in **d** as the phase boundaries. The possible Fermi surface structure (shown by schematics) for each phase is inferred by the FFT analysis of quantum oscillations, assuming that spin-valley flavors are not mixed. The superconducting region is marked by dashed lines. **f, i**, d$V_{xx}$/d$I_{dc}$ versus $I_{dc}$ of the hole- (**f**) and electron-doped superconductivity (**i**) at various $D$. **g, j**, $I_c$ (upper panel) and $T_c$ (lower panel) versus $D$ of the hole- (**g**) and electron-doped superconductivity (**j**). **h, k**, Temperature dependence of $R_{xx}$ versus $n$ on the hole-doped side at $D = 1.1$ V/nm (**h**) and on the electron-doped side at $D = -1.64$ V/nm (**k**). The insets show the $R_{xx}$ versus $T$ curves at the optimal doping, where $T_c$ reaches its highest value.

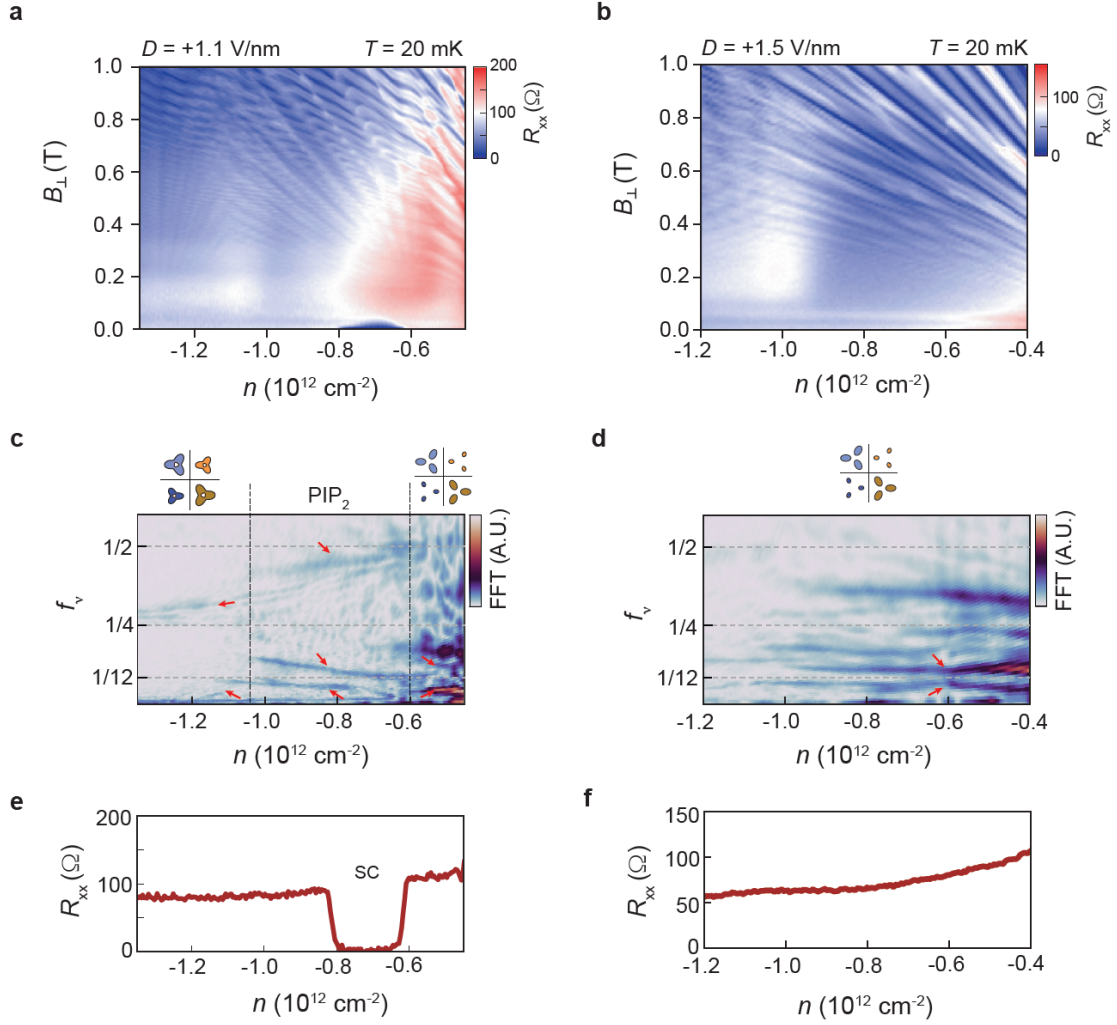

**Fig. 2. Fermi surface analysis of the hole-doped superconductivity. a, b**, $R_{xx}$ versus $n$ and $B_\perp$ at $D = 1.1$ V/nm (**a**) and 1.5 V/nm (**b**) on the hole-doped side. **c, d**, FFT of $R_{xx}$ ($1/B_\perp$) versus $n$ and $f_v$ at $D = 1.1$ V/nm (**c**) and 1.5 V/nm (**d**) on the hole-doped side. The FFT analysis in **c** and **d** is performed based on the $R_{xx}$ data within $0.2$ T $< B_\perp < 1$ T, respectively. The schematic Fermi surface structures for different phases are also shown in **c** and **d**. The frequency peaks of the FFT for different phases are highlighted by red arrows. **e, f**, $R_{xx}$ versus $n$ at $B = 0$ T measured at $D = 1.1$ V/nm (**e**) and 1.5 V/nm (**f**) on the hole-doped side. At $D = 1.1$ V/nm, a superconducting state is observed within the PIP$_2$ phase near the trigonal warping phase with the Ising SOC-induced spin splitting.

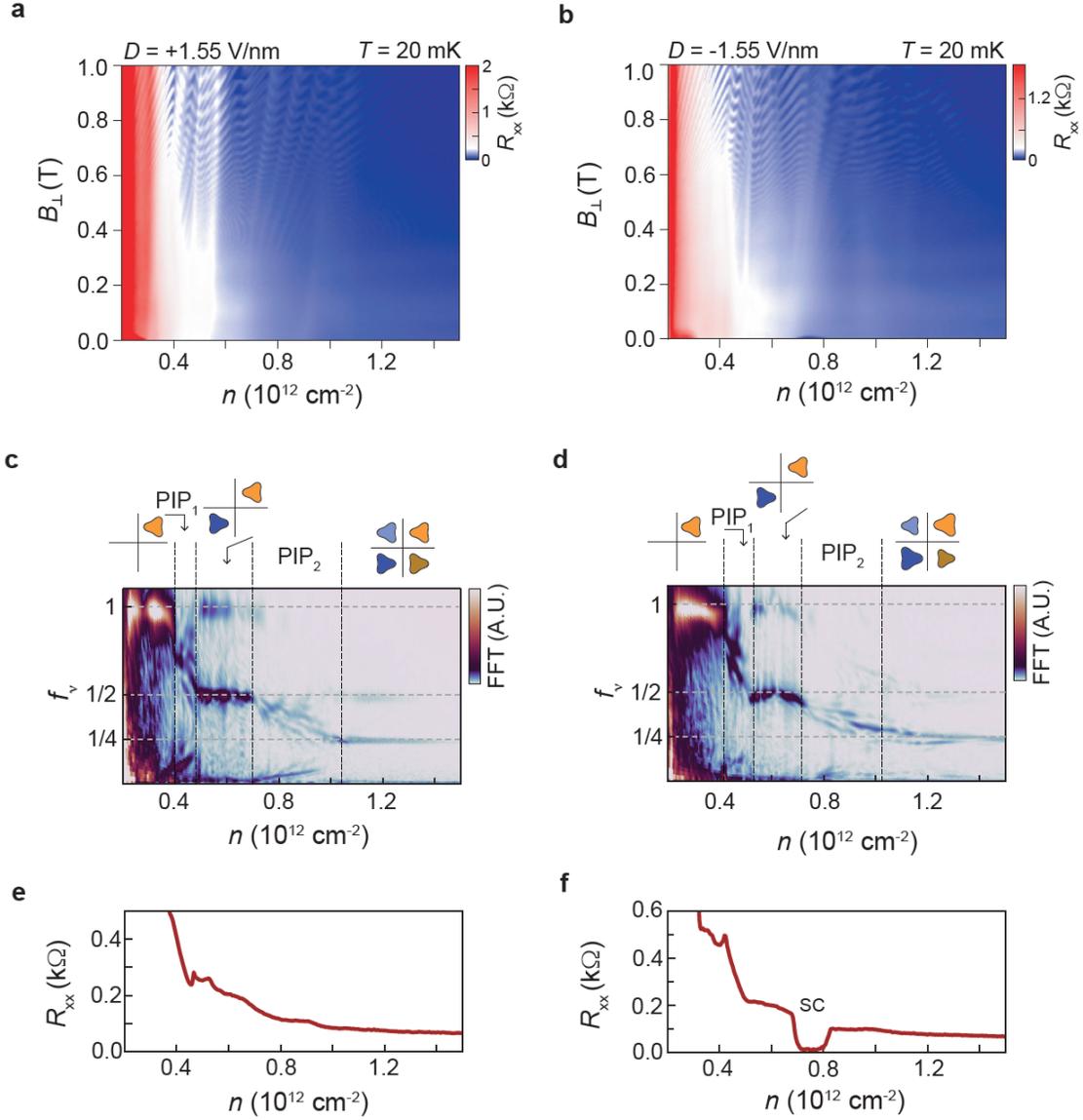

**Fig. 3. Fermi surface analysis of the electron-doped superconductivity. a, b**, $R_{xx}$ versus $n$ and $B_\perp$ at $D = 1.55$ V/nm (**a**) and -1.55 V/nm (**b**) on the electron-doped side. **c, d**, FFT of $R_{xx}$ ($1/B_\perp$) versus $n$ and $f_v$ at $D = 1.55$ V/nm (**c**) and -1.55 V/nm (**d**) on the electron-doped side. The FFT analysis in **c** and **d** is performed based on the $R_{xx}$ data within $0.2$ T $< B_\perp < 1$ T in **a** and **b**, respectively. The schematic Fermi surface structures for the different phases are also shown in **c** and **d**. **e, f**, $R_{xx}$ versus $n$ at $B = 0$ T at $D = 1.55$ V/nm (**e**) and -1.55 V/nm (**f**) on the electron-doped side. Electron-doped superconductivity can only be observed at negative $D$.

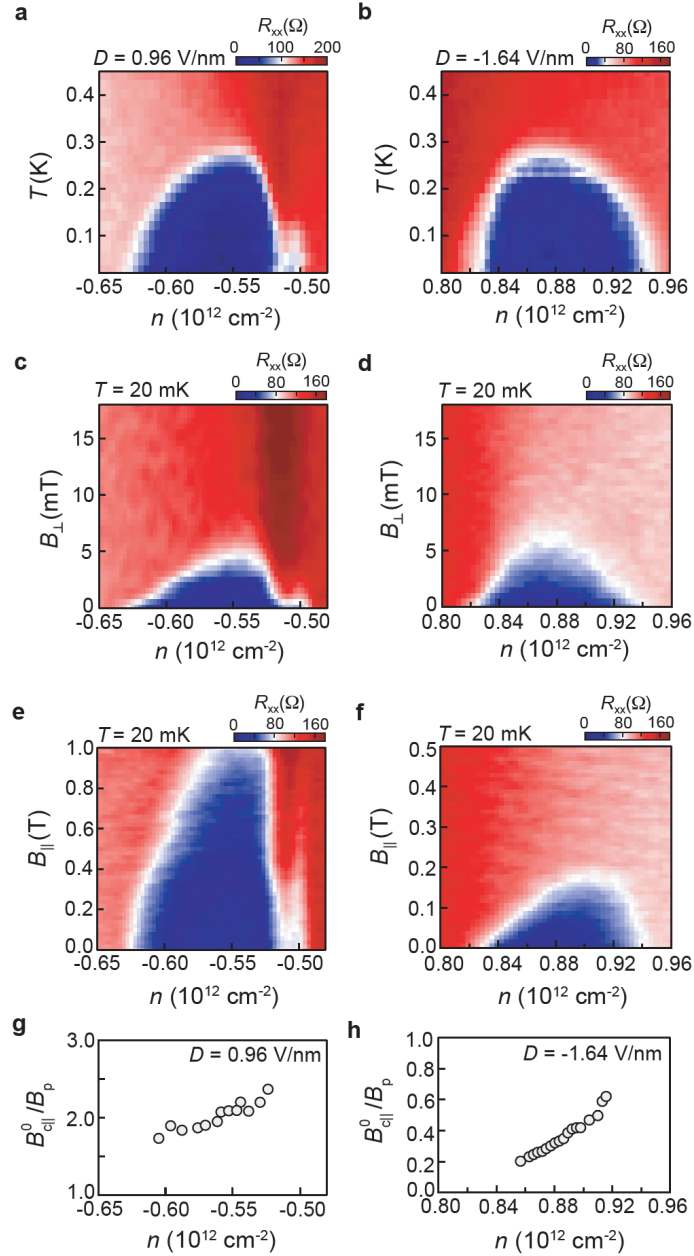

**Fig. 4. The in-plane magnetic field dependence of both hole-doped and electron-doped superconductivity. a**, **b**, $R_{xx}$ as a function of $T$ and $n$ at $D = 0.96$ V/nm for the hole-doped superconducting dome (**a**), and $D = -1.64$ V/nm for the electron-doped superconducting dome (**b**), respectively. **c**, **d**, $R_{xx}$ as a function of $B_\perp$ and $n$ at $D = 0.96$ V/nm for the hole-doped superconducting dome (**c**), and $D = -1.64$ V/nm for the electron-doped superconducting dome (**d**), respectively. Both the superconducting $T_c$ and $B_{c\perp}$ are similar at these two specific $D$ values. **e**, **f**, $R_{xx}$ as a function of $B_\parallel$ and $n$ at $D = 0.96$ V/nm (**e**) and $D = -1.64$ V/nm (**f**), respectively. The hole-doped superconductivity at the optimal doping could still survive at $B_\parallel$ up to 1 T, while the entire electron-doped superconducting dome at $D = -1.64$ V/nm is completely suppressed under a small $B_\parallel \sim 0.2$ T. **g**, **h**, Pauli violation ratio $B_{c\parallel}^0/B_p$ versus $n$ at $D = 0.96$ V/nm (**g**) and $D = -1.64$ V/nm (**h**), respectively. The hole-doped superconductivity violates the Pauli limit, while electron-doped superconductivity obeys the Pauli limit. Moreover, $B_{c\parallel}^0/B_p$ in both hole- and electron-doped superconductivity exhibit density dependent behaviors.

**Extended Data Figures**

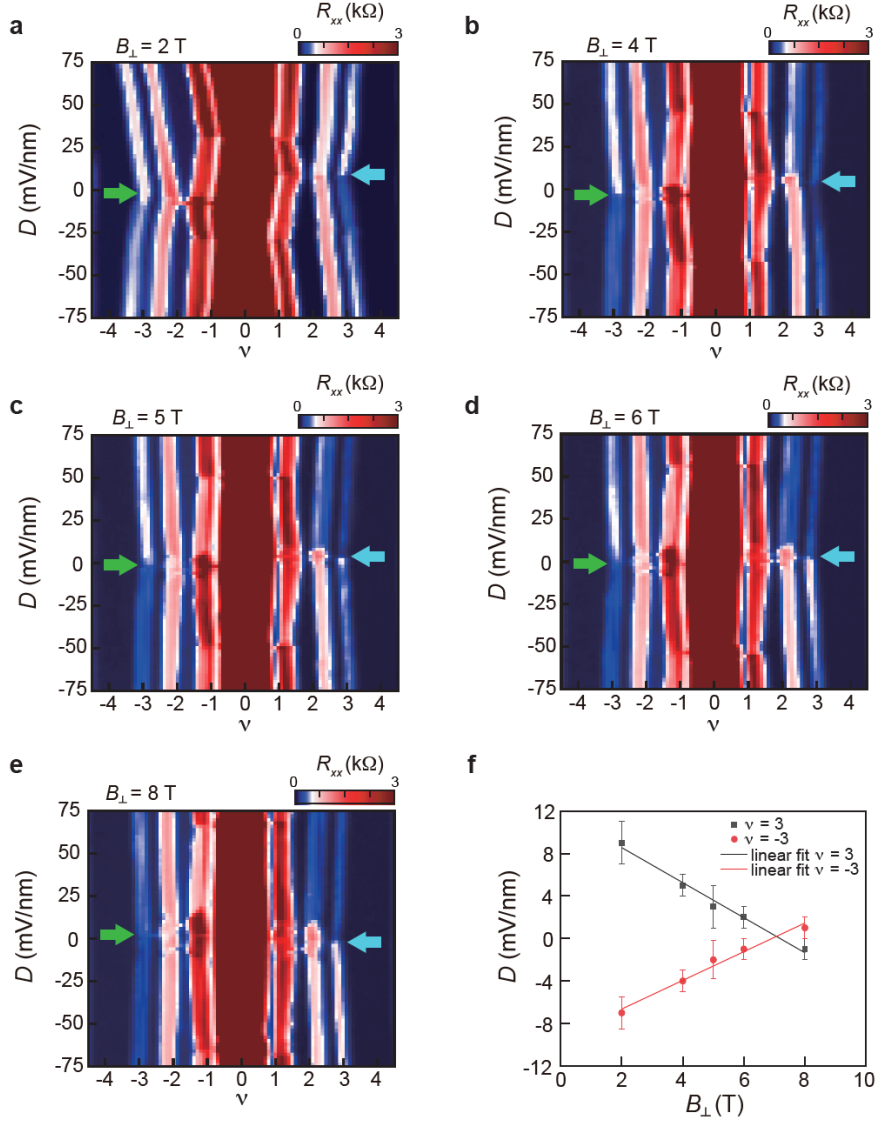

**Extended Data Fig. 1. Estimate the strength of Ising SOC from the transition at quantum hall state |ν| = 3. a-e**, $R_{xx}$ as a function of $D$ and Landau level filling factors ν at $B_\perp = 2$ T (**a**), $B_\perp = 4$ T (**b**), $B_\perp = 5$ T (**c**), $B_\perp = 6$ T (**d**) and $B_\perp = 8$ T (**e**). The blue and green arrows in each panel mark the orbital transitions of quantum hall state |ν| = 3. **f**, The $D$ extracted from the transition at |ν| = 3 in **a-e** as a function of $B_\perp$, and the red and black lines are fits to the data, respectively. The strength of Ising SOC could be estimated from the crossing point of the two fitting lines where the out-of-plane Zeeman energy $E_z$ compensates the energy split $\lambda_I$ due to Ising SOC[20,21,31]. According to $\lambda_I = 2E_z = 2g\mu_B B_{SOC}$, where $B_{SOC}$ is the perpendicular magnetic field where the two fitting lines intersect, the strength of Ising SOC $\lambda_I$ in our device is estimated to be about 1.7 meV.

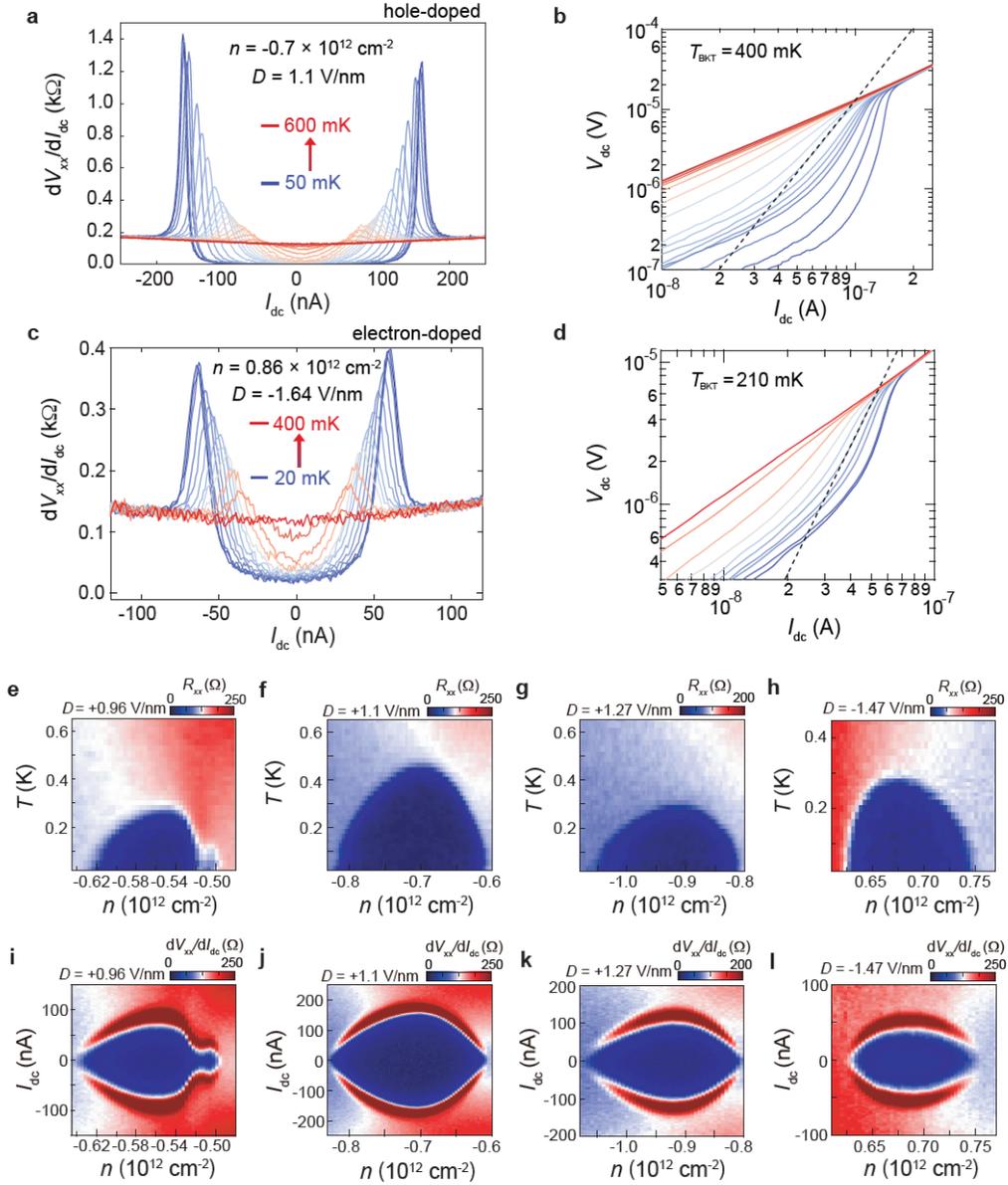

**Extended Data Fig. 2. Estimation of $T_{BKT}$, density dependent $T_c$ and $I_c$ at various $D$. a, c**, $dV_{xx}/dI_{dc}$ at the optimal doping as a function of $I_{dc}$ measured at various temperatures at $D = +1.1$ V/nm (**a**) and $D = -1.64$ V/nm (**c**) for the hole- and electron-doped superconductivity, respectively. An ac modulation current of 2 nA is used for the differential resistance measurements. **b, d**, The nonlinear voltage-current ($V_{dc}$-$I_{dc}$) curves of **a** and **c**. The dashed line is a power law fit of $V \propto I^3$, yielding $T_{BKT} = 400$ mK (**b**) and $T_{BKT} = 210$ mK (**d**) for the hole-doped and electron-doped superconductivity, respectively. **e-g**, $R_{xx}$ as a function of $n$ and $T$ for hole-doped superconducting domes at $D = 0.96$ V/nm (**e**), $D = 1.1$ V/nm (**f**) and $D = 1.27$ V/nm (**g**), respectively. **h**, $R_{xx}$ as a function of $n$ and $T$ for electron-doped superconducting dome at $D = -1.47$ V/nm. **i-l**, The measured differential resistance $dV_{xx}/dI_{dc}$ at $T = 20$ mK as a function of $n$ and dc bias current $I_{dc}$ for superconducting domes at (**i**) $D = 0.96$ V/nm, (**j**) $D = 1.1$ V/nm, (**k**) $D = 1.27$ V/nm and (**l**) $D = -1.47$ V/nm, respectively. A competing resistive phase intersecting the hole-doped superconducting dome reported previously[20] is evident at $D = 0.96$ V/nm and eventually diminishes with further increasing $D$. The data shown in **i-l** and the data shown in **e-h** were taken during separate rounds of measurements conducted in different dilution refrigerators. We found that the width of the superconducting dome in density can vary slightly between different rounds of measurements.

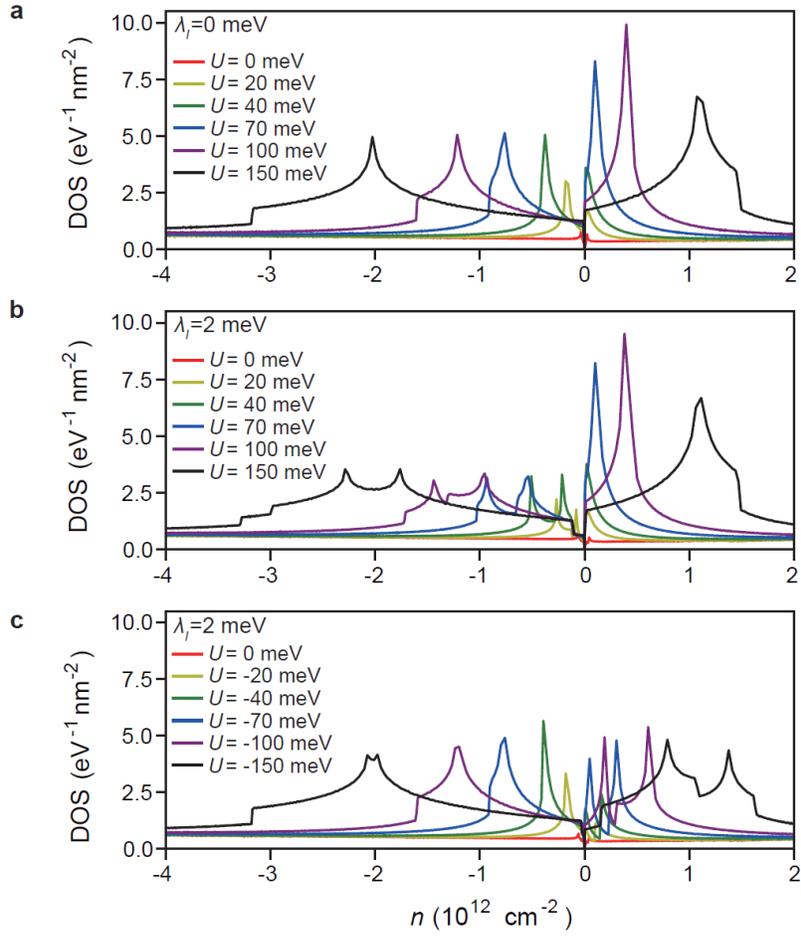

**Extended Data Fig. 3. DOS calculation. a-c** Total density of state (DOS) in BBG as a function of doping density $n$ without (**a**) and with (**b-c**) the Ising SOC term for different values of the layer potential difference $U$. At positive (negative) $U$, hole wavefunctions (electron wavefunctions) concentrate at the top graphene layer which is closer to the WSe$_2$ layer, so the proximity-induced Ising SOC is only notable in the VB (CB).

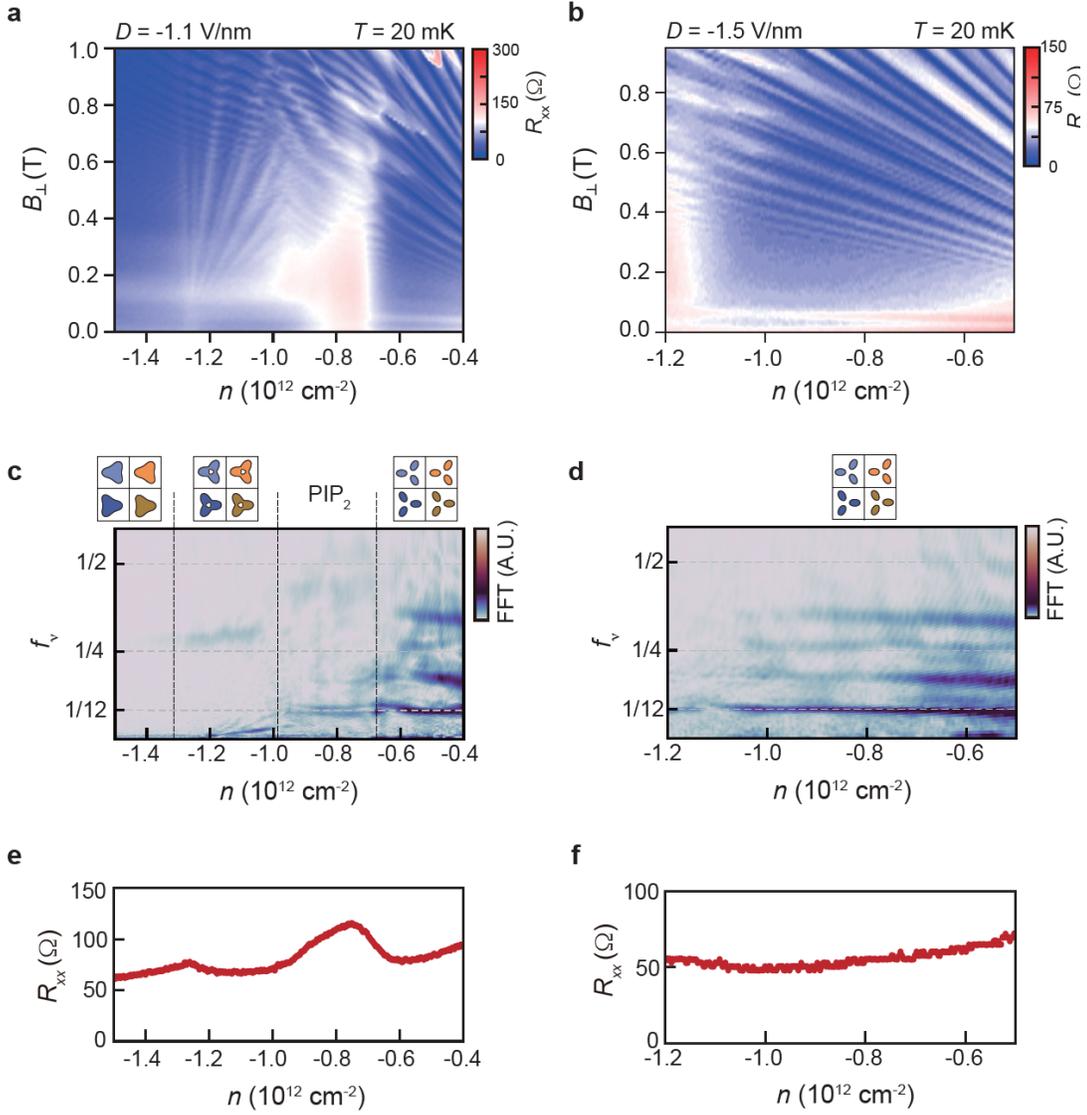

**Extended Data Fig. 4. Fermi surface analysis of the hole-doped BBG/WSe$_2$ at negative $D$ fields. a, b**, $R_{xx}$ versus $n$ and $B_\perp$ at $D$ = -1.1 V/nm (**a**) and -1.5 V/nm (**b**) on the hole-doping side. **c, d**, FFT of $R_{xx}$ ($1/B_\perp$) versus $n$ and $f_v$ at $D$ = -1.1 V/nm (**c**) and -1.5 V/nm (**d**) on the hole-doping side. The FFT analysis in **c** and **d** is performed based on the $R_{xx}$ data within 0.2 T < $B_\perp$ < 1 T in **a** and **b**, respectively. No SOC induced FFT peak splitting can be identified at negative $D$-fields on the hole-doping side. The schematic Fermi surface structures for different phases are also shown in **c** and **d**. **e, f**, $R_{xx}$ versus $n$ at $B$ = 0 T at $D$ = -1.1 V/nm (**e**) and -1.5 V/nm (**f**) on the hole-doping side. In the PIP$_2$ phase at $D$ = -1.1 V/nm, instead of superconductivity, a resistive state emerges.

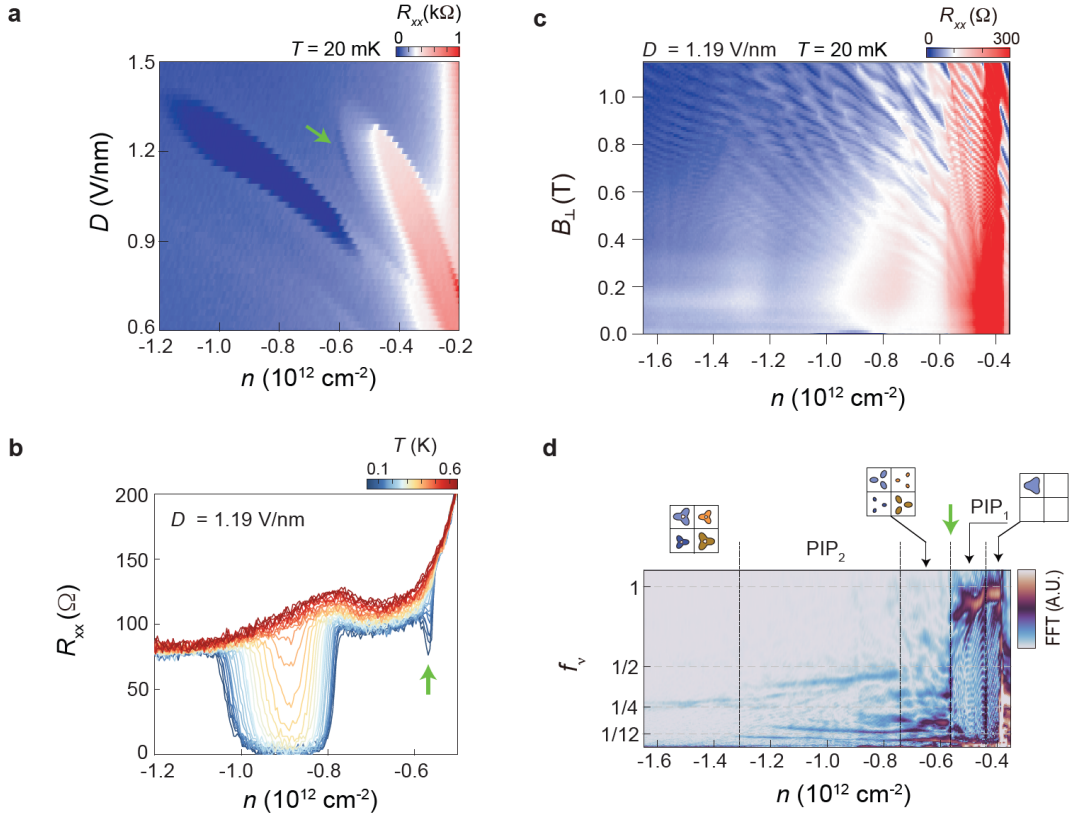

**Extended Data Fig. 5. Fermi surface analysis of the hole-doped BBG/WSe$_2$ at $D$ = 1.19 V/nm. a**, $R_{xx}$-$D$-$n$ map of hole-doped BBG/WSe$_2$ within a narrower $n$, $D$ range. Apart from the superconducting region described in the main text, another region with reduced $R_{xx}$ emerges at lower hole doping, within $D$-field range ~ 1.1 - 1.3 V/nm (marked by the green arrow). **b**, Temperature dependence of $R_{xx}$ versus $n$ on the hole-doping side at $D$ = 1.19 V/nm. The additional resistance dip at ~ $0.56 \times 10^{12}$ cm$^{-2}$ can be observed. Such resistance dip may indicate the developing of another superconducting dome, which may need further studies in higher quality devices or at lower temperatures. **c**, $R_{xx}$ versus $n$ and $B_\perp$ at $D$ = 1.19 V/nm on the hole-doping side. **d**, FFT of $R_{xx}$ ($1/B_\perp$) versus $n$ and $f_v$ at $D$ = 1.19V/nm on the hole-doping side. The FFT analysis is performed based on the $R_{xx}$ data within 0.2 T < $B_\perp$ < 1.2 T. A spin- and valley-polarized state with $f_v$ = 1 emerges at $n$ ~ -0.4 to -0.45 × 10$^{12}$ cm$^{-2}$. With increasing hole density, the FFT peak becomes less than 1 and new FFT peaks emerge at very low frequencies. These FFT features indicate a partially isospin polarized phase with one majority and multiple minority Fermi pockets (denoted as PIP$_1$ phase). Further increasing hole densities, the PIP$_1$ phase transits into the trigonal warping phase with the Ising SOC-induced spin splitting ($f_v^{(1)}$ > 1/12 and $f_v^{(2)}$ < 1/12) until $n$ ~ -0.75 × 10$^{12}$ cm$^{-2}$. The observed additional $R_{xx}$ dip locates in between of the PIP$_1$ phase and the trigonal warping phase, as indicated by the green arrow. Similar to $D$ = 1.1 V/nm shown in Fig. 2, the superconducting normal state is within the PIP$_2$ phase, corresponding to a partial isospin-polarized phase with two major Fermi pockets and multiple minor Fermi pockets. Further increasing hole doping beyond the PIP$_2$ phase, the system evolves into a state with four annular Fermi surfaces, which is evident by two FFT frequency peaks satisfying $f_v^{(1)}$ - $f_v^{(2)}$ = 1/4.

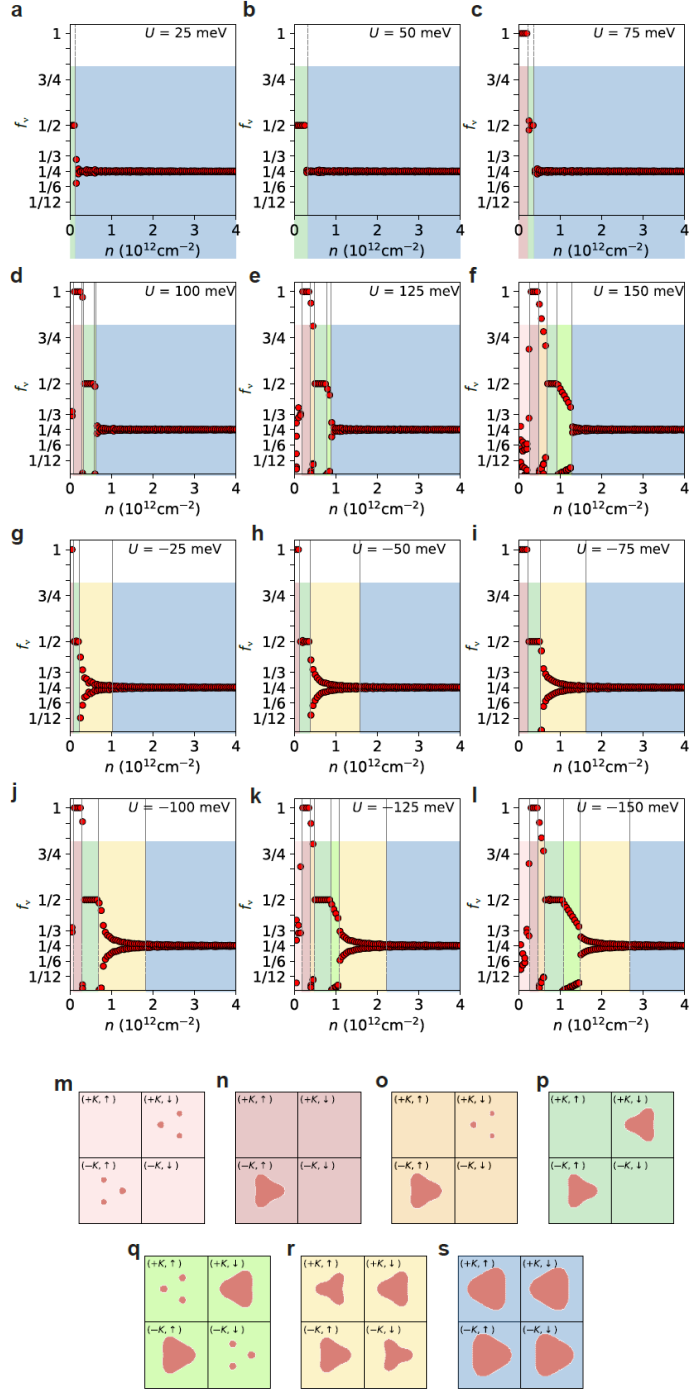

**Extended Data Fig. 6. The calculation of Fermi surface structure on the electron-doped side. a-l** Theoretically calculated normalized quantum oscillation frequencies $f_v$ as a function of $n$ for different values of $U$. We first calculate the mean-field ground state (considering both symmetric and symmetry-breaking states) at a given $n$ and $U$, and then $f_v$ is calculated by the fraction $S_i/S$, where $S_i$ is area of the ith Fermi pocket and $S = (2\pi)^2|n|$. The background colors distinguish different patterns of $f_v$. The results are presented for electron doping ($n > 0$). $U$ is positive in **a-f** and negative in **g-l**. The Ising SOC coupling strength $\lambda_I$ is taken to be 2 meV in the calculation. **m-s** Representative Fermi surfaces for different regimes in **l**. Electron densities in **m-s** are $n = (0.1, 0.4, 0.5, 0.8, 1.2, 1.6, 3) \times 10^{12}$ cm$^{-2}$, respectively.

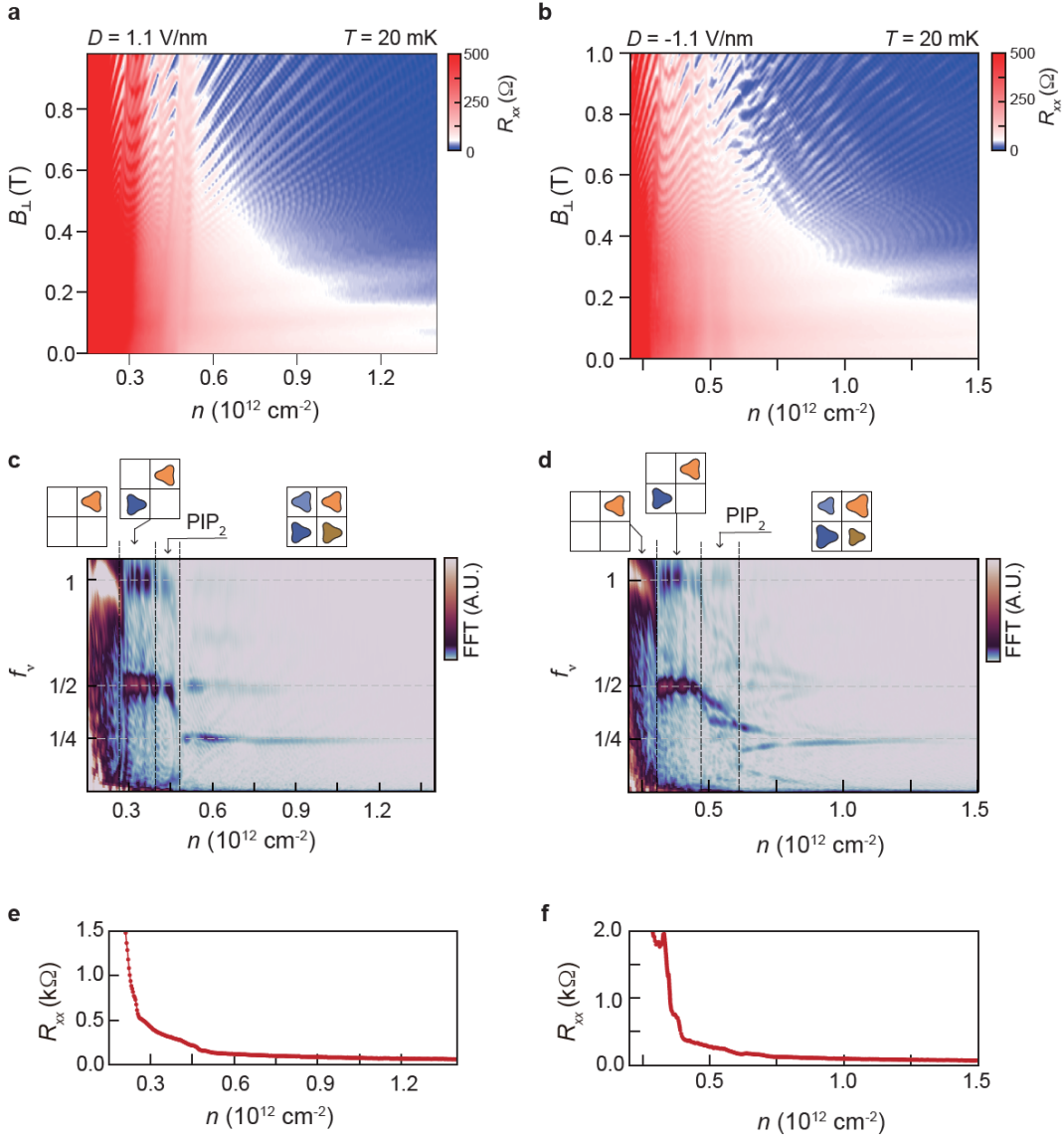

**Extended Data Fig. 7. Fermi surface analysis at $D = \pm 1.1$ V/nm on the electron-doping side. a, b**, $R_{xx}$ versus $n$ and $B_\perp$ at $D = 1.1$ V/nm (**a**) and -1.1 V/nm (**b**) on the electron-doping side. **c, d**, FFT of $R_{xx}$ ($1/B_\perp$) versus $n$ and $f_v$ at $D = 1.1$ V/nm (**c**) and -1.1 V/nm (**d**) on the electron-doping side. The FFT analysis in **c** and **d** is performed based on the $R_{xx}$ data within $0.1$ T $< B_\perp < 1$ T in **a** and **b**, respectively. The schematic Fermi surface structures for different phases are also shown in **c** and **d**. Compared to larger $D$ values (Fig. 3 and Extended Data Fig. 8), the PIP$_1$ phase is absent, and the electron density range of the PIP$_2$ phase become much narrower at $D = \pm 1.1$ V/nm. **e, f**, $R_{xx}$ versus $n$ at $B = 0$ T at $D = 1.1$ V/nm (**e**) and -1.1 V/nm (**f**) on the electron-doping side. Although the flavor-symmetry-breaking phases still exist, the superconductivity is absent at $D = -1.1$ V/nm.

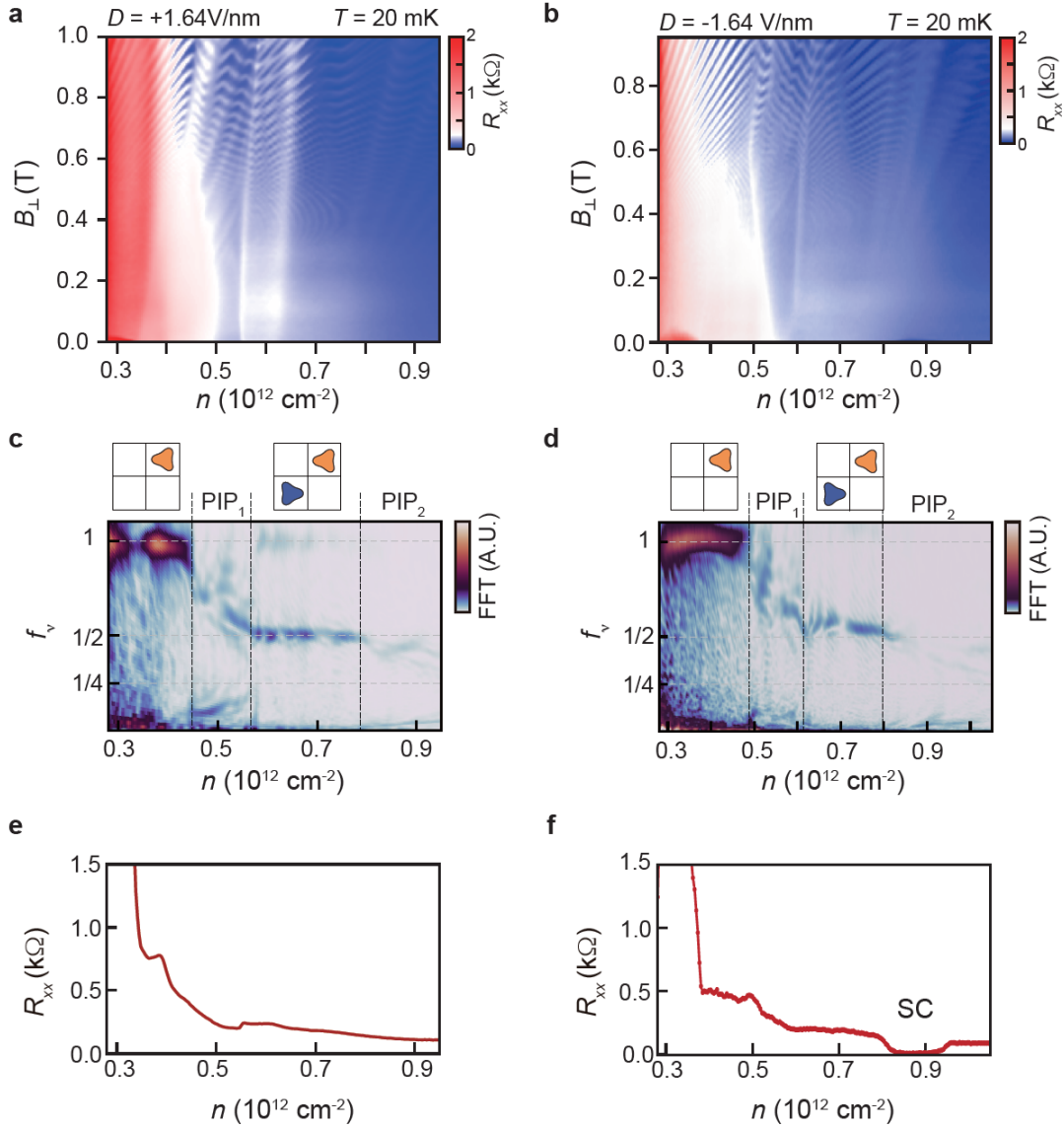

**Extended Data Fig. 8. Fermi surface analysis at $D = \pm1.64$ V/nm on the electron-doping side. a**, **b**, $R_{xx}$ versus $n$ and $B_\perp$ at $D = 1.64$ V/nm (**a**) and -1.64 V/nm (**b**) on the electron-doping side. **c**, **d**, FFT of $R_{xx}$ ($1/B_\perp$) versus $n$ and $f_v$ at $D = 1.64$ V/nm (**c**) and -1.64 V/nm (**d**) on the electron-doping side. The FFT analysis in **c** and **d** is performed based on the $R_{xx}$ data within 0.2 T < $B_\perp$ < 1 T in **a** and **b**, respectively. The schematic Fermi surface structures for different phases are also shown in **c** and **d**. **e**, **f**, $R_{xx}$ versus $n$ at $B = 0$ T at $D = 1.64$ V/nm (**e**) and -1.64 V/nm (**f**) on the electron-doping side. Electron-doped superconductivity can be only observed at negative $D$. The main results closely resemble those observed at $D = \pm1.55$ V/nm, as illustrated in Fig. 3.

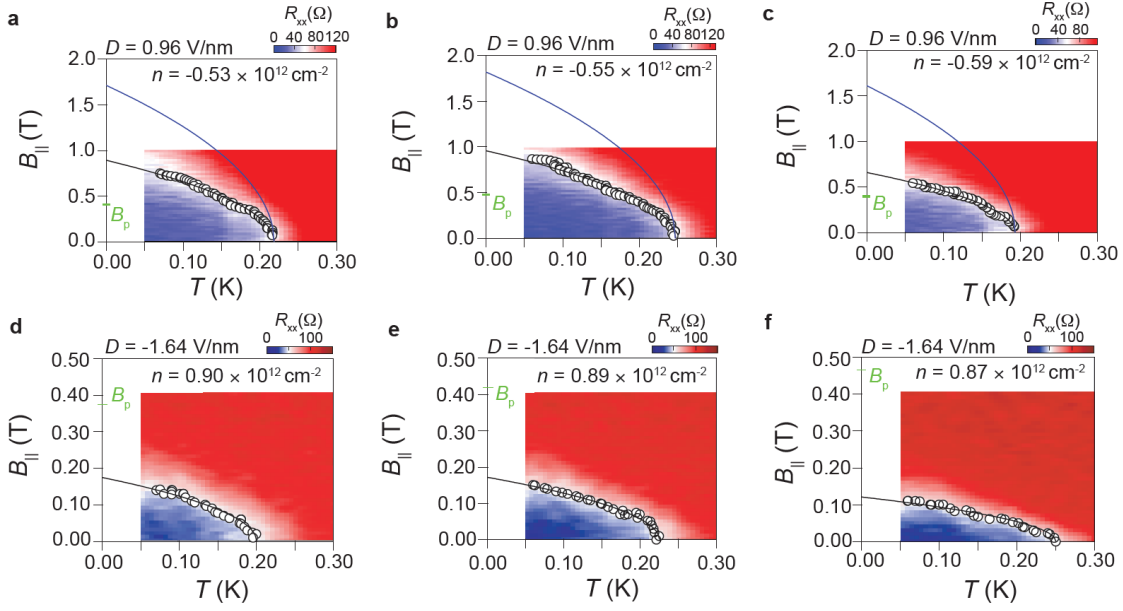

**Extended Data Fig. 9. Determination of the in-plane critical magnetic field at the zero-temperature limit $B^0_{c\parallel}$. a-c**, $R_{xx}$ as a function of $T$ and $B_\parallel$ at $n = -0.53 \times 10^{12}$ cm$^{-2}$ (**a**), $n = -0.55 \times 10^{12}$ cm$^{-2}$ (**b**), and $n = -0.59 \times 10^{12}$ cm$^{-2}$ (**c**) for $D = 0.96$ V/nm. **d-f**, $R_{xx}$ as a function of $T$ and $B_\parallel$ at $n = 0.9 \times 10^{12}$ cm$^{-2}$ (**d**), $n = 0.89 \times 10^{12}$ cm$^{-2}$ (**e**) and $n = 0.87 \times 10^{12}$ cm$^{-2}$ (**f**) for $D = -1.64$ V/nm. The opaque circles in each panel depict the critical in-plane magnetic field $B_{c\parallel}$ as a function of $T$, where the $B_{c\parallel}$ is defined as the field where $R_{xx}$ is 50% of the normal state resistance. The data points in each panel are fitted well by the phenomenological relation $T/T_c^0 = 1 - (B_{c\parallel}/B^0_{c\parallel})^2$. The green markers indicate the Pauli-limit field $B_P$. Blue lines are plotted based on the formula $T/T_c^0 = 1 - (B_{c\parallel}^2/B_P B_{SOC})$ for an Ising superconductor, where $B_{SOC}$ is obtained from our measurement shown in Extended Data Fig. 1. It can be seen that, even for the hole-doped superconductivity, the measured $B_\parallel$ is still smaller than the values expected for an Ising superconductor. Such discrepancy may depend on multiple details, including the Fermi surface shape, the Rashba SOC, the spin Zeeman effect, and the orbital effect of $B_\parallel$. As a general trend, the Ising SOC enhances PVR, while additional Rashba SOC and orbital effect from $B_\parallel$ suppresses PVR. Therefore, the value of PVR becomes a quantitative problem given these competing effects. On the other hand, quantitative estimation of quantities such as Rashba SOC, and orbital g-factor of the in-plane magnetic field is a nontrivial task, since they all have a small energy scale and are all subjected to renormalization by the electron Coulomb interaction. This makes it challenging to theoretically estimate the value of PVR. Nevertheless, the hole-doped superconductivity clearly violates the Pauli paramagnetic limit, consistent with previous studies[20,21]. However, the limited resilience to $B_\parallel$ observed in electron-doped superconductivity is more puzzling, as a comparable Ising SOC effect is evident in the CB at negative $D$ fields based on the FFT analysis of quantum oscillations.

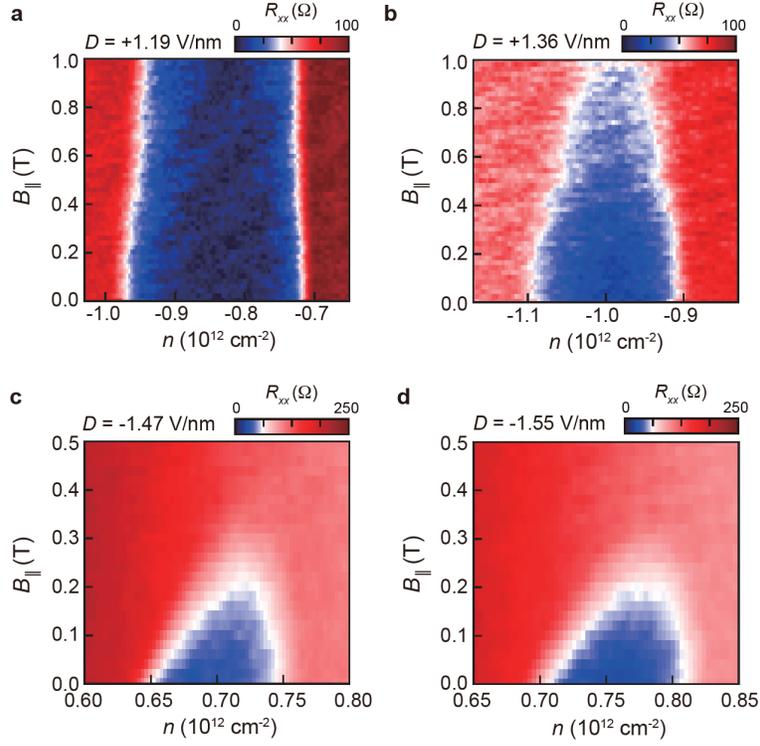

**Extended Data Fig. 10. More data about the in-plane magnetic field dependence of superconducting states**. **a-d**, $R_{xx}$ as a function of $n$ and $B_\parallel$ for hole-doped superconducting domes at $D$ = 1.19 V/nm (**a**), $D$ = 1.35 V/nm (**b**), and for electron-doped superconducting domes at $D$ = -1.47 V/nm (**c**) and $D$ = -1.55 V/nm (**d**), measured at $T$ = 20 mK. The superconducting dome width in $n$ at $D$ = 1.19 V/nm is almost unchanged under $B_\parallel$ = 1 T. At $D$ = 1.35 V/nm, the hole-doped superconductivity around -1 × 10$^{12}$ cm$^{-2}$ could still survive under $B_\parallel$ = 1 T. The highest in-plane magnetic field applied is limited to 1 T due to the magnet limitation of the refrigerator used for the measurement. The Pauli violation ratio $B^0_{c\parallel}/B_p$ at the optimal doping should be significantly larger than 1.4 in **a**, and ~ 2.1 in **b**. On the contrary, the electron-doped superconductivity in **c** and **d** is readily suppressed under a small applied $B_\parallel$ (about 0.2 – 0.3 T). The $B^0_{c\parallel}/B_p$ at the optimal doping for **c** and **d** is about 0.31 and 0.25, respectively, significantly below the Pauli paramagnetic limit.

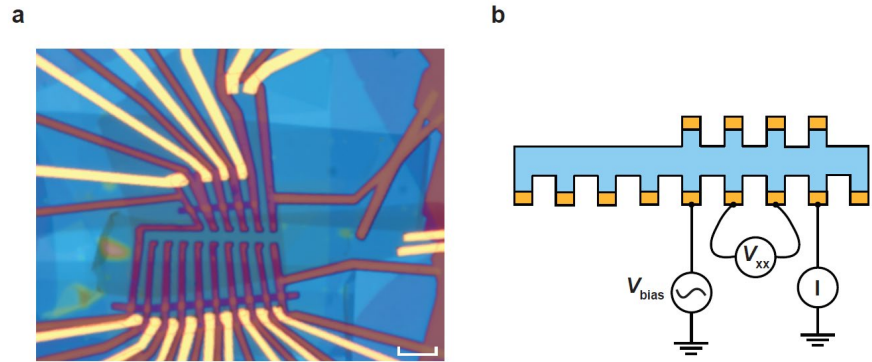

**Extended Data Fig. 11. Device image and the measurement configuration. a**, Optical image of the BBG/WSe$_2$ heterostructure device. The device is shaped into a hall bar geometry and the hall bar channel is fabricated in a bubble-free region. The scale bar is 5 μm. **b**, The schematic of the hall bar device in **a**, along with the illustration of the transport measurement configuration.

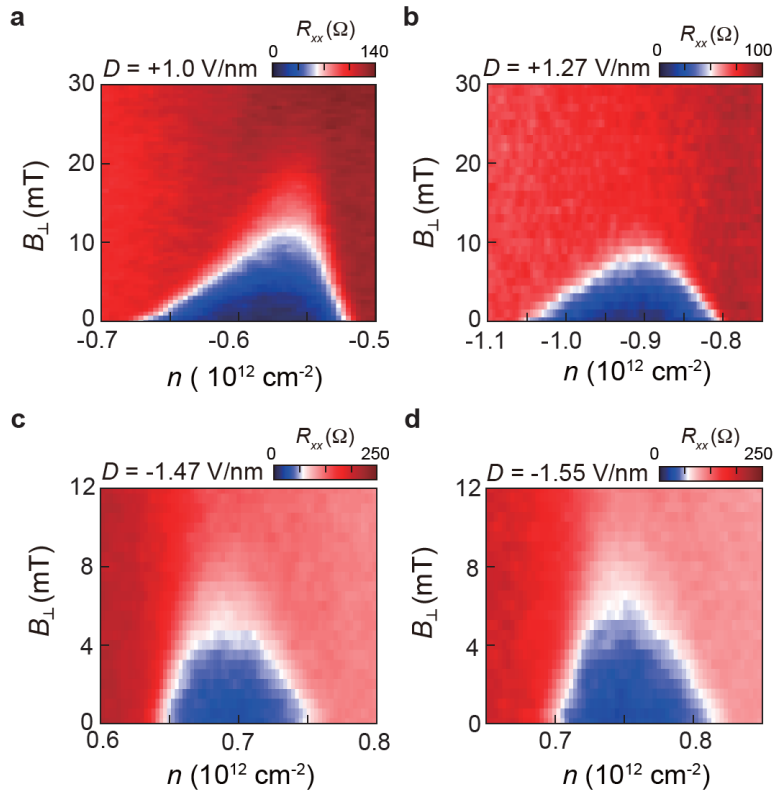

**Extended Data Fig. 12. The perpendicular magnetic field $B_\perp$ dependence of the hole- and electron-doped superconducting states. a-d**, $R_{xx}$ as a function of $n$ and $B_\perp$ measured at $T = 20$ mK with $D = 1.0$ V/nm (**a**), 1.27 V/nm (**b**) for hole-doped superconducting domes, and $D = -1.47$ V/nm (**c**), and -1.55 V/nm (**d**) for electron-doped superconducting domes, respectively. The critical perpendicular magnetic fields $B_{c\perp}$ for the hole- and electron-doped superconductivity are comparable, which range from about 5 mT to 15 mT.